\documentstyle[aps]{revtex}
\twocolumn

\begin{document}
\title{Comment on ``Maximum likelihood reconstruction of CP maps",
quant-ph/0009104 }

\author{Jarom\'{\i}r Fiur\'{a}\v{s}ek and Zden\v{e}k Hradil}

\address{Department of Optics, Palack\'{y} University, 17. listopadu 50,
772 07 Olomouc, Czech Republic}

\date{\today}
\maketitle

\begin{abstract}
The  treatment proposed by Sacchi quant-ph/0009104 does not
represent correct solution, since the necessary conditions on CP
maps  are not guaranteed.
\end{abstract}

Maximum--likelihood (Max-Lik) principle finds wide variety of
applications in quantum theory due to its ability to incorporate
necessary conditions as  constraints.  Max-Lik has been devised for
reconstruction of a generic quantum state  keeping the positive
semidefiniteness in Ref. \cite{Hradil}. Remarkably, it is not only
an estimation, but a genuine  generalized measurement
\cite{mereni}. In his recent papers Sacchi
\cite{preprint1,preprint2} applied Max-Lik to the reconstruction of
CP maps using the numerical algorithm of downhill-simplex method
\cite{downhill}. However, the proposed treatment does not represent
the correct Max-Lik solution consistent with quantum theory.

A trace preserving CP map is a linear map from operators in Hilbert
space ${\cal H}$ to operators in  ${\cal K}$. The mathematical
formulation  of CP map is  expressed by the relations (3-5) of the
paper \cite{preprint1}. The necessary conditions allowing physical
interpretation  are given by  relations (6-8). This may be treated
as a condition analogous to normalization of a density matrix
${\rm Tr} \varrho =1$ in quantum state reconstruction.
 However, in the case of CP map such a condition is given  by the
relation  (7) of \cite{preprint1}
\begin{eqnarray}
 S \ge 0, \quad {\rm Tr}_{\cal K} S = 1_{\cal H}. \label{rovnost}
\end{eqnarray}
 The condition (\ref{rovnost}) effectively represents $N^2$
 conditions, $N= {\rm dim}{\cal H}.$ 
They all  are correctly taken into account in Eq. (19)
of Ref. \cite{preprint1} using the Lagrange multiplieres
$\mu_{ij}$
\begin{equation}
{\cal L}_{\rm eff}[S]={\cal L}[S]-{\rm Tr}_{\cal H} [\mu ({\rm
Tr}_{\cal K}[S])].
\label{logl}
\end{equation}
Here $ {\cal L}[S]$ denotes the log-likelihood given for example by
the relation (2) of the paper \cite{preprint1}.
Unfortunately, the problem of finding the maximum
of  ${\cal L}_{\rm eff}[S]$ under the constraints (\ref{rovnost})
has not been solved. Using an excuse
that ``multipliers cannot be easily inferred'' the maximization has
been done under  ``looser constraint''
${\rm Tr} S = N$ only.
In this way,  Sacchi has fixed   the matrix of Lagrange
multipliers as $ \mu=(K/ N) 1_{\cal H}$, $K= {\rm dim}{\cal K}$.
He was probably inspired by the numerical Max-Lik solution for
quantum state estimation \cite{downhill}, where the
semipositiveness and trace normalization represent the only
constraints of quantum theory. However, this analogy is misleading
and improper in the case of CP maps allowing also nonphysical
solutions. Effectively, only a single condition instead of $N^2$
conditions is imposed in this case. As a consequence the
reconstruction  does not meet necessary conditions, namely the
relation ${\rm Tr}_{\cal K} S = 1_{\cal H}$.
This is obvious from the numerical
results, which, in fact, may serve as a counterexample. The
necessary conditions $S_p(1,1)+ S_p(3,3)=1$ and $S_p(2,2)+
S_p(4,4)=1$ are reproduced by the Table 1 as $0.995163231$ and
$1.006669754$. In spite of the author's claim that ``the estimated
values compare very well with the theoretical ones" the result does
not correspond to any CP map. For example, the former relation
means that the input state $|0\rangle $ is by such a ``device"
transformed into a state, where the total probability to appear on
the output at the states $|0\rangle $ and $|1\rangle $ is only
$0.995.$ The probability is therefore not conserved. Hence the
condition (\ref{rovnost}) is  as important as the semipositiveness
itself.

 Frankly, the proposed method hardly exhibits any significant
advantage in comparison to recently used linear reconstruction
methods. Linear approach is feasible for any dimension and all the
necessary conditions are also fulfilled ``approximately."  Linear
treatment corresponds to the maximization of the likelihood (2) in
\cite{preprint1} without any constraint. ``Max-Lik" reconstruction
of  Sacchi \cite{preprint1} imposes semipositiveness and single
constraint. This can be hardly considered as a significant
difference since $N^2$ constraints must be taken into account in
full quantum treatment.

 Max-Lik reconstruction of CP maps can be formulated correctly
\cite{Mirek}. The analogy between quantum state and CP map
reconstructions is established  on more sophisticated level than
assumed in Ref. \cite{preprint1}. As shown in Ref. \cite{Mirek}
solution for  multiplieres and CP  map may be obtained using an
iterative algorithm. In the case of single qubit, the solution may
be  found even using numerical downhill simplex method, provided
that the effective number of 12 parameters is used. Remarkably, the
extremal equation has the form of closure relation for probability
valued operator measure.  The Max-Lik reconstruction can be
therefore interpreted as a genuine quantum measurement in the same
sense as in the case of quantum state reconstruction \cite{mereni}.

\end{document}